\documentclass[12pt,preprint]{aastex}
 \begin{document}
 \title{ The Solar Heavy Element Abundances: II. Constraints from Stellar Atmospheres.}

\author{M.H. Pinsonneault$^1$ \& Franck Delahaye $^{1,2}$ \\
 1 Department of Astronomy, The Ohio State University, Columbus OH 43210 USA \\
 2 LUTH, (UMR 8102 associ\'ee au CNRS et \`a l'Universit\'e Paris 7),
   Observatoire de Paris, F-92195 Meudon, France.}

\begin{abstract}
 
Estimates of the bulk metal abundance of the Sun derived from the latest generation of model atmospheres are significantly lower than the earlier standard values.  In Paper I we demonstrated that a low solar metallicity is inconsistent with helioseismology if the quoted errors in the atmospheres models (of order 0.05 dex) are correct.  In this paper we undertake a critical analysis of the solar metallicity and its uncertainty from a model atmospheres perspective, focusing on CNO.  We argue that the non-LTE corrections for abundances derived from atomic features are overestimated in the recent abundance studies, while systematic errors in the absolute abundances are underestimated.  If we adopt the internal consistency between different indicators as a measure of goodness of fit, we obtain intermediate abundances [C/H] = 8.44 +/- 0.06, [N/H] = 7.96 +/- 0.10 and [O/H] = 8.75 +/- 0.08.  The errors are too large to conclude that there is a solar abundance problem, and permit both the high and low scales.  However, the center-to-limb continuum flux variations predicted in the simulations appear to be inconsistent with solar data, which would favor the traditional thermal structure and lead to high CNO abundances of (8.52, 7.96, 8.80) close to the seismic scale.  We argue that further empirical tests of non-LTE corrections and the thermal structure are required for precise absolute abundances.  The implications for beryllium depletion and possible sources of error in the numerical simulations are discussed.
\end{abstract}


\section{Introduction}

The uncertainty in the absolute chemical composition of stars is the limiting factor in our ability to do high precision stellar astrophysics.  Traditionally, we have had to rely on a small database of fundamental stellar parameters such as mass, distance, and radius.  However, current and upcoming space missions promise a wealth of astrometric and photometric data.  Large surveys undertaken primarily for other purposes (microlensing, planet searches and cosmology) have discovered thousands of eclipsing binaries, yielding numerous precise mass estimates.  The rapidly developing field of optical interferometry has also permitted a growing number of direct radius estimates.  Asteroseismology is also growing in importance, and missions such as COROT promise a wealth of detailed information on the pulsational properties of solar-like stars.

Our stellar interiors models have become highly sophisticated and successful when compared with observational diagnostics.  In particular, the resolution of the solar neutrino problem in favor of the solar model predictions and the agreement between theoretical predictions and helioseismic data are both encouraging signs.  The combination of better observations and theory has opened the prospect of a new era of precision stellar astrophysics, which could have broad consequences for diverse subfields of astronomy.

Stellar atmospheres theory has traditionally employed a series of approximations when deriving abundances.  Classical models assume an ad hoc turbulent velocity field adjusted to yield abundances independent of excitation potential and line strength.  Convection is usually treated in an approximate fashion, with the mixing length theory.  Horizontal temperature fluctuations (granulation) are not included.  The models also typically assume that the molecular and atomic levels are described by local thermodynamic equilibrium (LTE), e.g. by the local temperature alone.  The compilations of solar abundances used for theoretical solar models \citep{ag1989,gn1993,gs1998} employed model atmospheres with approximations at the level described above.  The mean thermal structure employed was semi-empirical \citep[hereafter HM74]{hm1974}.

When these approximations are relaxed, different conclusions about the abundances are obtained.  Departures from LTE are expected at a modest level for solar conditions, and have been investigated by a number of authors \citep[for example][]{c1986,sh1990,k1993,stb2001,w2001}.  Numerical simulations of convection have matured to the level where they can be used to predict velocity fields, temperature fluctuations, and changes in the mean thermal structure of the upper atmosphere \citep{sn1998}.  Abundances derived from these simulations yield a very different pattern, which has been developed in a series of papers \citep{pla2001,pla2002,asplund2000a,asplund2000b,agsak2004,agsak2005}; papers by Lodders (2003) and \citet{ags2005} summarize the revised abundance scale.  

The net effect is in the sense of systematically lower metal abundances.  The downward revisions for the heavier elements (e.g. Fe and Si) are small, while the claimed reduction in the abundances of lighter species (especially CNO) is more dramatic.  Models employing different treatments of granulation and non-LTE corrections \citep{h2001,sh2002} predict smaller abundance reductions.  The central temperature predicted by interiors models is sensitive to the abundances of the heavier elements, but not the lighter ones.  As a result, the new abundance scale does not disturb the agreement between interiors models and observational data for purposes such as the mass-luminosity relationship and solar neutrino fluxes.  However, the inferred solar sound speed profile, and the radii of interiors models, is sensitive to the bulk metallicity. Serious problems have emerged when comparing interiors models with the revised abundance scale.  These discrepancies are evidence for problems in our understanding of stellar interiors, stellar atmospheres, or both.  In Paper I \citep{dp2006} we investigated the errors in solar abundances predicted by the combination of stellar interiors models and helioseismic data.  In this paper we examine the uncertainties in the abundance predictions from stellar atmospheres theory.  We begin with a brief summary of the results from Paper I, and follow with a discussion of the motivation and main results from the revised stellar atmospheres models.  In Section 2, we perform a critical analysis of the precision of the solar CNO abundances and discuss the implications for Be.  We demonstrate in that section that the errors in the abundances are larger than previously estimated, and that there is evidence that the ``best'' current solar CNO abundances are intermediate between the new and old scales, with errors permitting both.  We discuss the implications of our finding and future tests in Section 3. In particular, we argue that inconsistencies between the solar thermal structure and that predicted by the simulations would favor a higher abundance scale closer to the seismic value and discuss uncertainties in the numerical convection simulations.

\subsection{Constraints from Helioseismology}

Helioseismology provides two powerful constraints on the solar composition: diagnostics of the internal solar temperature gradient and diagnostics of the equation of state.  Inversions of the observed solar pulsation frequencies yield accurate measures of the sound speed as a function of depth.  In turn, the gradient in the sound speed can be directly tied to the temperature gradient.  Since the temperature gradient is related to the opacity, and thus the composition, information on the solar abundances is encoded in the seismic data for the radiative interior.  One can even obtain meaningful constraints on the solar age from the helium abundance profile deduced in the deep interior.

In addition to the vector information on the sound speed profile,
there are also precise scalar quantities that can be extracted.  The
thermal structure at the base of the solar convection zone is nearly
adiabatic, while the temperature gradient in the interior is
radiative.  As discussed in Paper I, Section 2.2 the resulting
discontinuity in $\nabla$ generates a distinct signal that can be used
to precisely localize the base of the solar convection zone
\citep[$R_{cz}= 0.7133 \pm 0.0005 R_{sun}$][]{ba2004}.  Seismology also sets strict limits on convective overshooting \citep[$< 0.05 H_P$,][]{cdmt1995}.  The depth of the solar convection zone is sensitive to the light metal abundances in the Sun but insensitive to most of the other uncertainties in solar interiors models (see Paper I for a detailed error budget).  Ionization also induces a depression in the adiabatic temperature gradient, and the absolute abundances of the species in question can be inferred from the magnitude of the perturbation in the surface convection zone.  An extremely precise surface helium abundance can be deduced from this effect ($Y_{surf}=0.2483 \pm 0.0046$, see Paper I, Section 2.3 for the sources used in this estimate).  More recently, \citet{ab2006} have demonstrated that the ionization signal of metals in the convection zone can be detected in the seismic data, leading to a bulk metallicity $Z=0.017 \pm 0.002$.  Because the majority of the solar metals are in the form of CNO, this is primarily a constraint on their abundance.  In principle one might be able to use this technique to solve for individual heavy element abundances by fitting the strength of distinct ionization stages.  However, it is not yet clear that there is sufficient spatial resolution in the seismic data to permit such a detailed analysis.

In Paper I we demonstrated that the combination of the surface convection zone depth and surface helium abundance constraints was a powerful diagnostic of the solar heavy element abundances.  The surface helium abundance is tied to the initial solar helium abundance with a correction for gravitational settling.  The initial helium is sensitive to the central opacity and the abundances of the heavier metals (especially iron).  The convection zone depth is sensitive to the opacity at temperatures ~ 2 million K, where bound-free opacity from light metals (CNONe) is an important contributor.  The most significant new finding in Paper I was that the combination of the two scalar constraints could be used to rule out some abundance combinations with high statistical significance.  The detailed sound speed profile adds additional information; we found that models consistent with the scalar constraints could be constructed with low oxygen and very high neon, but such models exhibited substantial sound speed deviations relative to solar data in the deep interior.

The inferred solar oxygen and iron abundances ($[O/H]=8.86 \pm 0.05, [Fe/H]= 7.50 \pm 0.05$) are consistent with the \citet{gs1998} absolute abundances, but strongly inconsistent with the new abundance scale within its quoted errors \citep{l2003,ags2005}.  Although there are potentially positive chemical evolution consequences for the revised abundance scale \citep{turck2004}, it is not easy to generate interiors models that are consistent with both seismology and the low abundance scale.  The most commonly cited possible explanations on the interiors side (high neon, enhanced gravitational settling, and errors in the high temperature radiative opacities) are all strongly disfavored.  As previously mentioned, solutions with high neon degrade agreement with the sound speed profile, and are also problematic from a stellar atmospheres perspective \citep{s2005}.  An increase in the degree of gravitational settling increases the convection zone depth but decreases surface helium, trading improved agreement with one diagnostic for worse agreement in another.  Enhanced differential settling of metals with respect to helium is inconsistent with the underlying physics and would have to be extreme \citep{gwc2005}.  Three independent quantum mechanical calculations yield extremely similar Rosseland mean opacities at the temperatures of interest for the base of the solar convection zone.  As discussed in Paper I, both the atomic physics and equation of state are relatively simple in this regime, and the concordance between different calculations is thus not a surprise.  Physical processes neglected in classical stellar models (such as rotational mixing and radiative acceleration) can be independently constrained by other data, and in any case they would tend to induce higher rather than lower surface abundances.  The scalar constraints are insensitive to the other theoretical ingredients in standard solar models (e.g. convection theory, surface boundary conditions, low temperature opacities, equation of state, and nuclear reaction rates).  The considerations above indicate that it is extremely challenging to reconcile a low solar metal abundance with current stellar interiors models and seismic data.  This does not imply that a metal-poor Sun is impossible, but it certainly motivates an investigation of the uncertainty in the atmospheres models used to derive the abundances.

\subsection{Model Atmosphere Ingredients}

The new solar abundance estimates are derived from a variety of changes in the model atmospheres.  Changes in oscillator strengths and equivalent widths of spectral lines contribute for some diagnostics, and as discussed below we largely concur with the revised values.  The magnitude of non-LTE corrections depends on the atomic model and the relative importance of photo-excitation and collisions on the level populations in the model atmosphere.  The comprehensive re-examinations of the solar oxygen \citep[hereafter AGSAK04]{agsak2004} and carbon \citep[hereafter AGSAB05]{agsab2005} adopted a particular model for NLTE corrections, and we assess its accuracy and uncertainties by comparison with limb darkening data and other published calculations.  In \citet{pla2001,pla2002} the abundances derived from forbidden lines have been reduced by the application of blending corrections; AGSAK04 and AGSAB05 used the revised equivalent widths for C and O abundance studies.  Both the uncertainty in these corrections and their central value is incorporated in our error analysis.

Three other coupled changes in the atmosphere models are the treatment of convective velocity fields (``macro/microturbulence''), horizontal temperature fluctuations (granulation), and the impact of convective overshooting on the mean thermal stratification.  All of these features are derived from numerical convection simulations; a good discussion can be found in \citet{sn1998}.  Their combined impact can be deduced by the comparison of the results in the 3D case in the published studies with the results from the semi-empirical 1D HM74 thermal structure and the theoretical 1D MARCS models.  Since the initial 3D model atmosphere is derived with physics similar to that in the MARCS code, the impact of the convection treatment can be indirectly inferred by comparing MARCS and 3D abundances.  A comparison of HM74 with MARCS and 3D is a measure of the impact of different choices of the thermal structure.

The numerical convection simulations predict line profiles in excellent agreement with the data for iron and silicon lines \citep{asplund2000a,asplund2000b}.  There are some trends with excitation potential that may be related to NLTE corrections \citep{stb2001} and issues with the initial generation of simulations when compared with line profiles in the outer solar photosphere \citep{scott2006}.  The concordance of the predicted amplitude of horizontal temperature fluctuations with the solar granulation pattern is encouraging \citep{sn1998, asplund2000a}.  The validity of the steep solar temperature gradient in the simulations is not as clear; there is an apparent conflict between the simulations and the mean thermal structure of the solar atmosphere, as well as the degree of convective penetration for the upper atmospheric layers \citep{ay2006}.  Since these effects are all tied together in the abundance studies, we focus on the agreement between different diagnostics of abundance as a valuable test of the precision of the results obtained from different atmospheric models.

\section{The Solar CNOBe Abundances}

Revised solar abundances have been derived for a number of species.
In this paper we focus on CNO for several reasons.  First, these are
the elements where the difference in abundance is largest, and they
are also the cases where there is the largest variety of distinct
abundance indicators.  The difference between interiors and
atmospheres based abundances for heavier species, such as Fe and Si,
are not statistically significant.  Furthermore, details of the new
abundance estimates and internally consistent comparison with prior
work are published for only some of the heavier elements. The solar Be
abundance is an important diagnostic of mixing, and the photospheric
abundance is linked to the solar O (Balachandran \& Bell 1998).  We therefore also briefly discuss the implications of our result for Be. We begin with a description of our overall approach, and then follow with individual sections on oxygen, carbon, and nitrogen.  Our primary references for O, C, N are respectively AGSAK04, AGSAB05, and \citet{ags2005}.  We include other studies for external comparisons, and discuss newer results based on other models and diagnostics when available.	

\subsection{Overall Approach}

Our primary metric for the accuracy of the abundances derived from the different model atmospheres is the consistency of the estimates derived from distinct classes of indicators.  One might justifiably apply a different standard, noting for example the greater degree of sophistication in the input physics for the hydrodynamic simulations.  However, it is not clear that the ingredients that induce the abundance changes (such as changes in the temperature gradient) are actually a necessary consequence of these improvements in the model, as the absolute errors in the first-principles theoretical models have not been quantified.  We proceed in two steps.  In the first, we compute relative abundances and errors within a given assumed model atmosphere for the atomic and molecular features.  Since the abundances from atomic lines have smaller model-to-model differences (and thus smaller systematic errors), we adopt the atomic abundances for our base estimate.  The difference between atomic and molecular abundances is then used to infer which of the different atomic scales should be adopted for the central value, and the uncertainty in the differential scales is used as a measure of the systematic error arising from the choice of model atmospheres.

For the atomic line diagnostics, we include random errors from the dispersion of results from single permitted lines about the adopted mean.  For the forbidden lines, uncertainties in oscillator strengths and blending corrections become the dominant random error source.  We also include systematic errors (NLTE corrections and zero-point shifts in the average oscillator strength) by comparing the study values with external constraints and other published calculations.  This is important when comparing the atomic and molecular indicators, since NLTE corrections are usually included in the former but not the latter.  As a result, changes in the degree of NLTE corrections have a direct impact on goodness of fit.  When available, we adopt a weighted mean of the permitted and forbidden atomic indicators and the error in the mean when comparing with the molecular data.

For the molecular indicators, we include all measured lines of a given diagnostic and use the total dispersion (rather than the error in the mean) as a measure of goodness of fit.  We adopt a weighted mean of the various indicators for an average molecular abundance.  However, the error in the mean will understate the true uncertainty; it is frequently the case that the mean values for different molecular diagnostics differ by more than the dispersion within each indicator.  We therefore compute the dispersion of each indicator about the adopted mean and average these values to obtain an error in the molecular abundances.  One could alternately compute the dispersion in the molecular abundances and treat this as a measure of the systematic uncertainty, adding it to the error in the mean in quadrature; this procedure yields somewhat smaller errors.  Although the latter approach may be practical when there are numerous molecular probes available, we prefer the former method for situations like oxygen (where there are 2 values, and thus an unreliable estimate of the uncertainty in the mean).

We derive final abundance estimates for each species by comparing the mean abundances derived within each class of models for atomic and molecular indicators.  In the case of oxygen, the 3D and HM74 models exhibit comparable differences with opposite sign, while the MARCS models have an internally consistent intermediate abundance.  We therefore adopt a mean of the different derived oxygen abundances and an uncertainty from the scatter.  In the case of nitrogen, the HM74 abundances are preferred.  The case of carbon depends on the origin of systematic differences in abundances inferred from atomic features.  If the low zero point of AGSAB05 is adopted, the 3D models are the favored solution.  If the higher zero point of previous work is adopted, the situation is similar to that for oxygen.

\subsection{Oxygen Abundance Indicators}

AGSAK04 derived a low solar oxygen abundance ($8.66 \pm 0.05$) from four distinct indicators: atomic lines (forbidden and permitted) and two different classes of infrared molecular lines ((v,r) and (r,r)).  All four indicators had formerly been used to obtain higher absolute oxygen abundance (8.83 to 8.90).  
Abundance estimates from all indicators are reduced in the theoretical atmospheres that include substantial overshooting because lines become stronger for a fixed abundance in the presence of a steeper temperature gradient.  Molecular abundances are reduced more than atomic ones because the cooler atmospheric structure of the 3D hydro models changes the chemical equilibrium.   AGSAK04 included other effects that reduced the abundances derived from atomic features without impacting the molecular indicators: a combination of changes in oscillator strengths, the inclusion of blending features, and their claim of large non-LTE effects for the permitted lines.

In this section we discuss the uncertainties in each of these cases.  We advocate a substantial decrease in the magnitude of the NLTE corrections for the [O/H] derived from the permitted atomic oxygen lines, and an increased error in both the abundance derived from the forbidden line (from uncertainties in the oscillator strength) and the permitted lines (from uncertainties in the NLTE corrections).  We also derive an increased error from the [O/H] derived from the IR molecular lines from the internal scatter and trends with excitation potential, and argue that the correspondence between the trends in the MARCS and 3D models is evidence for errors in the common underlying model.  Our basic results are summarized in Table 1.  
In Table 1, the upper part of the table repeats the mean abundances and errors for the 4 different indicators presented in AGSAK04.  We present our revised estimates for the same three cases in the lower part.  The last three rows in each sub-table give the mean discrepancy between atomic and molecular indicators.
At the end of the section we synthesize this information to obtain our best estimate for the solar oxygen abundance.  In each of the following subsections, the error estimates derived are internal ones.  The differences between the results from the three classes of models are prima facia evidence that systematic errors are important.  The systematic errors are discussed in the final subsection. 

\subsubsection{Forbidden Oxygen Lines}

The forbidden oxygen line at 6300.3 \AA~ has traditionally yielded high oxygen abundances.  \citet{pla2001} argued that a nearby blended Ni line contributed significantly to the oxygen feature.  They treated the continuum level, log (gfNi), and the oxygen abundance as free parameters, but assumed that the line profiles as given from the simulations were exact.  The inclusion of the Ni feature induced a direct reduction of 0.13 dex in the inferred oxygen abundance.  In addition, the usage of a 3D model atmosphere structure led to a further reduction of 0.08 dex in the oxygen abundance to an estimated [O/H] = 8.69 +/- 0.05.  In AGSAK04 another forbidden line at 6363.7 \AA~ was considered as a second indicator.  The authors reduced the equivalent width by 0.5 m\AA~ for an estimated contribution from a blended CN feature to obtain an oxygen equivalent width of 1.4 m\AA~, also implying a low abundance.  We begin our analysis by noting that abundances derived from blended features are usually treated with great caution.  The most conservative procedure is to ignore the blending feature and treat the derived abundance as an upper limit.  When this is done for the two forbidden lines, the maximum abundance obtained for 3D, HM, and MARCS are (8.82, 8.8), (8.9, 8.88), (8.86, 8.84) respectively.  In the section that follows, we include the reduction in abundances from estimates of the blending contribution.

We adopt the AGSAK04 values for [O/H] derived from the 6300.3 \AA~ line, subject to the caution on the strength of the Ni blending feature below.  For the 6363.7 \AA~ line, \citet{m2004} argued that the (10,5) Q$_2$ 25.5 CN line is unblended in the solar spectrum and has the same oscillator strength as the feature blended with the forbidden line.  He derived a smaller correction for the blended CN line (0.35 m\AA~ rather than the 0.5 m\AA~ value used in AGSAK04), which we adopt here.  This leads to a modest 0.04 dex increase in the derived abundance from that line, which we also treat as an uncertainty in the [O/H] derived from this feature from the uncertainty in the contribution of CN to the blended feature.
The error analysis for a blended feature is more complex than the one that can be employed for an isolated line.  The derived [O/H] is sensitive to the continuum level, and an error component for this should be included; \citet{pla2001} estimate this uncertainty at 0.02 dex, which we adopt for both lines.  AGSAK04 adopted lower values for the oscillator strengths than those found in the NIST database, but their choice is well-supported by the improved atomic physics \citep[see][]{sz2000} .  However, the errors in individual theoretical log(gf) values are higher than those assigned in \citet{pla2001} and subsequent papers; we adopt 0.04 dex for individual lines.  \citet{pla2001} also estimated that uncertainties in the underlying equation of state induces errors of 0.02 dex.

Especially for the 6300 \AA~ line, the results depend heavily on the detailed line profiles, particularly in cases where the individual components cannot be directly disentangled.  \citet{pla2001} estimated uncertainties of 0.02 dex from the central wavelength of the Ni feature and a 0.04 dex uncertainty from the central wavelength of the [O I] line.  We treat these errors as representative of the uncertainties in the line profiles, and adopt them for both forbidden lines.

The treatment of the Ni line in the main forbidden line is more problematic.  In the initial study, the oscillator strength was highly uncertain.  \citet{pla2001} treated log (gf) for Ni as a free parameter.  The continuum level, $log (gf_{Ni})$, and [O/H] were treated as free parameters and the combination that produced the minimum $\chi^2$ was adopted.  However, \citet{j2003} have measured the oscillator strength of the Ni feature (log gf = -2.11), and the Ni abundance of the solar mixture is well-constrained by knowledge of the solar Si/Fe and the relative meteoritic abundances.  With the new gf value and [Ni/H]=6.25 the blending feature would be 0.23 dex stronger than the best $\chi^2$ value obtained in the 2001 paper.  AGSAK04 did not report the inferred strength of the blending feature in their fit, but the similarity in the absolute abundance suggests that it would be comparable.  We believe that it is no longer appropriate to treat this as a free parameter, and the same method should be used as is done for other blended features: namely, the strength of the Ni line should be held fixed (and varied within its uncertainty) while the free parameters are the oxygen abundance and continuum level.

The direct effect of increasing a blending contribution is usually to decrease the abundance, but the present case is more complicated.  For the quoted values in the AGSAK04 fit, the Ni line contributes 25\% of the total equivalent width of the line.  An increase of 0.23 dex in the strength of the feature would imply a total Ni contribution of 43\% of the blended equivalent width at a fixed continuum level; such a combination would be a poor fit to the line shape and would yield a reduction in the inferred oxygen of 0.12 dex.  An increase in the continuum level would be required to restore the agreement with the line profile, which would in turn lead to an increased total equivalent width.  The net impact on the derived oxygen is not obvious, and not necessarily in the negative direction.  \citet{r1998} invoked log gf=-1.95 for the Ni feature and obtained [O/H]=8.75 for the forbidden line when his oscillator strength for the forbidden line was adjusted to the same value as that employed by Allende Prieto et al.  In the absence of other information, we assign an additional error component of 0.04 dex for the strength of the Ni feature, slightly higher than the value advocated by \citet{m2004}.

Adding the errors in quadrature, we obtain an uncertainty of 0.078 per line (0.055 in the average) and abundances derived from the forbidden lines systematically 0.02 dex higher than those found in AGSAK04. 

\subsubsection{Permitted Atomic Oxygen Lines}

The permitted atomic oxygen lines have relatively high oscillator strengths, but very high excitation potentials.  The most commonly used features are the OI triplet at 7771.8, 7774.2, 7775.4 \AA~, with an excitation potential of 9.15 eV.  
AGSAK04 also considered three other atomic features (6158.1 \AA~, 8446.7 \AA~, and 9266 \AA~).  The primary reason for the low oxygen abundance inferred by AGSAK04 from the triplet is a large non-LTE correction.  Because these lines arise from such a high energy state, non-LTE effects must be included.  However, the quoted values of the non-LTE corrections in the literature vary drastically.  For the triplet, the average is -0.06 dex for \citet{h2001}, -0.22 to -0.28 for AGSAK04, and -0.16 dex for \citet{pafb2004}.  These variations can be partially traced to different assumptions about the importance of collisional excitation (as opposed to photoionization), but larger differences at the 0.10 dex level remain even for cases that make similar assumptions about hydrogen collisions.

AGSAK04 neglected hydrogen collisions in their estimate of the non-LTE effects.  They justified this by noting that for some well-studied lines, the classical \citet{d1968} formulism overestimates the collision rate.  However, an inspection of their Figure 6 indicates that the neglect of collisional effects in their model yields changes as a function of limb darkening that differ from the observed solar values.  This impression is confirmed by the more detailed study of \citet{pafb2004}, who found that models including hydrogen collisions ( their $S_H=1$ case) were a better fit to the solar data.  We therefore conclude that the non-LTE corrections in AGSAK04 are overestimated, which has a significant effect on the concordance of the different oxygen indicators.

AGSAK04 applied larger downward reductions to the HM model than to the 3D and MARCS models, which we do not believe to be justified. \citet{pafb2004} indicate that similar corrections are obtained for Kurucz and 3D models in their detailed study of the triplet as a function of limb darkening.  This is particularly important because the discordance between the oxygen derived for the 1D models from atomic and molecular lines was used as a primary argument for the superiority of the 3D models, and this discrepancy can be directly traced to the assignment of very large NLTE corrections to the HM model.  By very similar logic, the internal dispersion in abundance for the atomic lines in the 1D case arises from the assignment of large NLTE corrections to some of the lines; the internal agreement of the HM case is improved (and that of the 3D and MARCS cases degraded) with smaller NLTE effects. We illustrate this point in Figure 1.  For the range of NLTE corrections that we consider reasonable (shaded band) the internal dispersion for the 3D models exceeds that of either 1D model. This emphasizes the role of substantial NLTE corrections to the scientific conclusions of AGSAK04.

In order to quantify this effect, we normalized the non-LTE corrections in Table 3 of AGSAK04 to an average for the triplet of three values: 0.16 dex (the best fit from \citet{pafb2004} and 0.11 dex (the mean between Holweger 2001 and \citet{pafb2004}), and 0.06 dex \citep{h2001}.  We adopt the 0.11 dex level as our best case for reasons outlines below.  The NLTE corrections for the other lines were scaled by the linear ratio of the average NLTE corrections for the triplet and the target values.  We present revised abundance estimates for the permitted lines in AGSAK04 computed in this manner in Table 2.  In Table 2, the first set of values contains the LTE results.  The next three sets represent calculations where the NLTE corrections were normalized to obtain a mean triplet correction of 0.06 dex, 0.11 dex (our adopted mean), and 0.16 dex.  The final set of results includes the NLTE corrections originally applied in AGSAK04.  This procedure yields non-LTE [O/H] abundances of (8.69, 8.73, and 8.76) for 3D models, 1D HM74, and 1D MARCS respectively in the 0.16 dex case and (8.73, 8.76, 8.80) in the 0.11 dex case.

Since the NLTE corrections are significant for the triplet, the uncertainty in these corrections is a major ingredient in the error budget.  Even the reduced NLTE corrections of Allende Prieto et al. (2004) for the triplet are substantially larger than the corrections used by Holweger (2001), who found an average NLTE correction of -0.06 dex, 0.10 dex lower than the value reported by AGSAK04.  Surprisingly, none of the authors involved commented on the origin of the difference.  Holweger (2001) obtained average LTE and NLTE triplet abundances of 8.78 and 8.72 respectively; his NLTE abundance is close to that obtained for the two 1D models in AGSAK04.  From \citet{pafb2004}, the case with no hydrogen collisions was ruled out at the $3 \sigma$ level, and LTE models were ruled out with high confidence.  However, the authors did not consider whether even lower NLTE corrections than their $S_H = 1$ case would have provided improved fits to the data.  In cases such as this, we see no justification for simply adopting one NLTE correction (0.16 dex) over another published value (0.06 dex), and adopt the average of the two (0.11 dex).  We note that our central values are close to what we would infer if we simply took the triplet alone as an oxygen abundance indicator and assigned a 0.16 dex NLTE correction.

We used the dispersion in the abundances derived from individual lines as a base random error.  A change of 0.05 dex in the triplet NLTE correction yields an average change in [O/H] of 0.035 dex, which we include as an additional systematic error.  Finally, the log (gf) values from AGSAK04 are lower than previously published values by \citet{bhgvf1991}; we therefore add another 0.025 dex systematic error, following a similar error analysis by \citet{m2004}.  The net effect is a total error estimate (summarized in Table 1) of 0.064 to 0.074 dex.

\subsubsection{Infrared Molecular Lines}

The IR molecular oxygen lines have been the primary abundance indicator used in previous compilations of solar abundances \citep[e.g.][]{gs1998}, and 1D model atmospheres yield relatively high absolute oxygen abundances. The absolute abundances are a strong function of the thermal structure of the model atmospheres, and the different thermal structure of the 3D model of AGSAK04 yields much lower predicted abundances than the 1D models.  We note that \citet{h2001} discounted CNO abundances derived from molecular lines because of their high temperature sensitivity.

AGSAK04 considered two sets of OH lines: (v,r) and (r,r).  The abundances predicted as a function of excitation potential from the 3D hydro simulations are compared with those from the 1D Holweger-Mueller and MARCS codes in Figures 2 and 3.  On these figures we have also indicated the average atomic abundances for each class of models.  We note the presence of striking trends with excitation potential in the 3D and MARCS models for the (r,r) lines.  The correspondence between the MARCS and 3D trends indicates that the origin of these features is common to both models, which indicates that it is a feature of the base atmospheric treatment rather than being induced by the convection simulation.

In the context of 1D models, trends such as those seen in the 3D models would be interpreted as a problem with the thermal structure or the assumed microturbulence.  AGSAK04 noted this trend, and claimed that it could be removed by invoking an outer atmosphere structure even cooler than the one predicted by the simulations.  Figure 1 suggests an alternate explanation, namely that the thermal structure in the outer layers is closer to the hotter semi-empirical HM74 model.    We will return to this point when we consider more recent work on CO abundances in the outer solar atmosphere.

AGSAK04 discarded the (r,r) data for the weaker and stronger lines, in effect deriving an abundance from the valley in Figures 2 and 3.  There is no better justification for discarding the high than the low points in this figure; such a procedure is not required for the 1D models.  We therefore derived average abundances using all of the features for all three models and both indicators; the standard deviation about the mean is an indicator of the quality of the fit for the individual bands.  Our results are summarized in Table 1.  The average molecular abundances were obtained with a weighted mean, but a simple averaging of the errors underestimates the dispersion.  We therefore computed $\sigma$ for each band around the weighted molecular mean in Table 1 and averaged the (v,r) and (r,r) values to obtain the total error in the molecular abundances presented there.

\citet{m2004} considered a third molecular oxygen abundance indicator, and derived relative abundance patterns comparable to what AGSAK04 found.  We do not include this in Table 1 because it is not clear that systematic errors between model atmospheres codes can be properly accounted for in a differential analysis.  If we had included the \citet{m2004} the mean molecular abundances would have been minimally altered for 3D and HM.  The average would be reduced for MARCS and the internal error in the MARCS [O/H] would be dramatically increased.  This provides further evidence that there is an underlying issue in the thermal structure of the MARCS model.  Melendez also computed abundances with the same indicators as AGSAK04 for a Kurucz model atmosphere, and derived similar abundances as would be found for the HM74 model. \citet{ay2006} present evidence for a high solar oxygen derived from CO studies; we postpone a discussion of this interesting result to our conclusion, in the context of tests of the solar thermal structure.

\subsubsection{Oxygen Abundance and Error Analysis}

Our overall result from the reanalysis of the AGSAK04 oxygen indicators is that the abundances derived from the atomic indicators are systematically increased for all models.  In the original paper, the 3D abundance estimators were found to yield consistent abundances, while the 1D abundances from different methods were highly discordant.  This conclusion no longer holds when the reduced NLTE corrections inferred from limb darkening studies are employed.  In fact, one would obtain very similar conclusions to those presented in Table 1 from the triplet abundance of 8.72 presented in Allende Prieto et al. (2004) for the 3D model (e.g. the 0.16 dex case presented in Table 2).

Rather than simply adopting one model or another as correct, we interpret the difference between the HM74 and 3D abundances as evidence that the thermal structure of the Sun is intermediate between the two.  The difference between the atomic and molecular abundances is roughly equal in magnitude and opposite in sign between these models; the MARCS model yields an intermediate abundance where the two classes of indicators give the same abundance, but with a larger error.  The error in (Atomic-Molecular) is substantial; these differences are formally significant only at slightly more than $1 \sigma$.  We therefore argue that the mean of the derived atomic abundances (8.75) is a reasonable estimator of what one would obtain from a model with a thermal structure capable of reproducing the atomic and molecular data; one would obtain 8.74 from comparing HM and 3D, and 8.76 from MARCS alone (which is already internally consistent).  We have a random error of 0.05 dex for the mean atomic abundance, but this is insufficient for a total error because of the presence of strong systematic differences.  Adopting the consistency between atomic and molecular indicators as a measure of goodness of fit, $~1 \sigma$ deviations could make either the 3D model ([O/H]=8.68) or the HM model (8.80) consistent.  We treat this as a $1 \sigma = 0.06 $ systematic error, and note that it is comparable to the zero-point shift that we obtain for the atomic indicators relative to AGSAK04 estimated below.

We can also examine systematic errors by comparing the AGSAK04 values with abundance estimates by other authors.  These are most easily analyzed by comparing LTE abundances for the triplet.  Holweger (2001) derived an average LTE triplet abundance of 8.78 for his standard model and 8.85 for the alternate VAL model.  \citet{bhgvf1991} reported 8.84 for the triplet for a HM model and 8.78 for a MACKKL atmosphere. These should be compared with 8.89, 8.87, and 8.93 for the three LTE cases in AGSAK04.  We note that the 1D cases in AGSAK04 used the equivalent widths obtained with the line broadening of the 3D hydro models, so there will be differences between their results and those obtained with other 1D codes.  The average LTE abundance is 8.85, with $\sigma = 0.055 dex$.  Systematic differences at the 0.06 dex level are thus a reasonable estimate of the current state of the art for oxygen abundances when estimated with different techniques.

Adding systematic (0.06) and random (0.05) errors in quadrature, we obtain 8.75 +/- 0.08 as our final oxygen abundance estimate for the Sun.  This is less than $1.6 \sigma$ below the helioseismic abundance, and therefore we conclude that the existence of a solar oxygen problem has not been demonstrated with high statistical significance.

\subsection{Carbon Abundance Indicators}

The overall story for carbon follows a similar path to the changes in the
inferred oxygen abundance, and the comprehensive reanalysis of AGSAB05 for
carbon has a similar logical structure to the 2004 oxygen paper.  Although
both the carbon and oxygen are reduced, the C/O ratio is preserved.  A cool
outer solar atmosphere in the 3D models yields substantially reduced
abundances from molecular indicators, while blending features and non-LTE
effects reduce the carbon abundance inferred from atomic features.  What
distinguishes the carbon from the oxygen case is that the NLTE effects are
smaller, and as a result one might anticipate a smaller offset in the atomic
line abundances than the change in atomic oxygen abundance indicators.
However, this is not the actual published result; if anything, the derived
carbon from atomic features is lower than the molecular value for all of the
models presented in AGSAB05.  Furthermore, we will demonstrate that this
effect cannot be explained by any of the effects used to explain the
differences in the comparison with prior work given by AGSAB05.  Until the
origin of this difference is understood, we therefore have to consider two
different systematic sets of abundances for atomic features, and the best
choice of model hinges on which set is correct.  In this section, we discuss
the three classes of indicators (forbidden and allowed atomic lines, and molecular) in turn, and as for the oxygen abundance synthesize our final best estimate and error in the fourth subsection.  Our overall estimates are presented in Table 3.  The top set of values represents the original AGSAB05 values for the different indicators.  
The middle set is what we would obtain with the low AGSAB05 normalization of the atomic abundances, while the bottom set is what we obtain with the higher Biemont/Holweger normalization.

\subsubsection{Forbidden Lines}

The [CI] line at 8727 \AA~ was the subject of a detailed analysis by \citet{pla2002}.  They incorporated blending from a nearby Si feature to reduce the equivalent width attributable to carbon from 6.5 to 5.3 m\AA~, with a corresponding reduction in the inferred abundance.  They also employ a lower oscillator strength than previous studies. Unlike the case of oxygen, the carbon abundance derived from the forbidden line is almost as temperature sensitive as that derived from molecular features, so the atomic versus molecular diagnostic is less powerful for carbon than for oxygen. We adopt the AGSAB05 central values for our base case, but note that there may be explained systematics in the atomic carbon abundances in AGSAB05 which we discuss below.   We include their error estimate for uncertainties in the equation of state (0.02 dex), but assign a larger uncertainty to the atomic physics (0.04 dex) in accord with the quoted theoretical uncertainties.  Our principle reservation on the error budget is the uncertainty in the continuum level and the contribution to the equivalent width of the blend from the wing of the Si feature.  Their reduced $\chi^2$ permits only small deviations (of order 0.01 dex) in the derived carbon abundance, but the base model relies upon the assumption that the underlying velocity field is exact.  Although the overall agreement with Fe \citep{asplund2000a} and Si \citep{a2000} line profiles is good, it is not errorless.  We cannot evaluate this ingredient directly, but an estimate based upon the mean deviation observed in clean lines would seem to be a worthwhile exercise.  For the present, we therefore assign the same blending uncertainty of 0.03 dex adopted by \citet{pla2002} to obtain a total uncertainty of 0.054 dex.    

\subsubsection{Permitted Atomic Lines}

AGSAB05 considered a subset of the permitted atomic features used in previous solar abundance studies \citep[e.g.][]{bhgv1993,sh1990}.  \citet{sh1990} found that small NLTE corrections are required for CI, and that the strength of the correction depends on equivalent width.  They found an average of -0.05 dex; if restricted to the weaker lines included in AGSAB05, their average NLTE correction would be -0.02 dex.  AGSAB05 computed Non-LTE corrections for 1D models, and the MARCS corrections were applied to the 3D models.  Hydrogen collisions were not included in the NLTE corrections; this resulted in larger downward abundance revisions (an average of -0.08 to -0.09) than \citet{sh1990}.  AGSAB05 note that the case of carbon should be an analog of oxygen, and we concur.  As a result, we contend that the case with hydrogen collisions should be included in the base model.  Both sources indicate that including hydrogen collisions roughly halves the expected NLTE correction.  We therefore considered two cases for NLTE corrections: a maximum of half the AGSAB05 value (corresponding to their hydrogen collision case, average -0.04 dex) and a minimum of one quarter of the AGSAB05 value (corresponding to the SH90 case, average of -0.02 dex).  Our best value is the average between the two (a mean of -0.03 dex), and the error induced by uncertainties in NLTE corrections is 0.01 dex; adopting the AGSAB05 hydrogen collision case would only have changed our mean value by 0.01 dex.  We applied these proportional NLTE corrections to the AGSAB05 LTE results for their three classes of models (middle values, Table 3).  There is a small reduction in the dispersion (and mean trend with equivalent width) for the 1D models and a corresponding increased for both in the 3D model; none of these features, however, are drastic.  The internal dispersion in the permitted atomic abundances is of order 0.03 dex.

A more substantial issue emerges when we compare the AGSAB05 abundances with prior work, and this is true even for the LTE estimates.  The mean LTE abundance for the \citet{bhgv1993} sample is 8.56 for the lines in common with AGSAB05; this should be compared with a HM LTE value of 8.48 for the latter compilation.  This offset of 0.08 dex is comparable to the average difference between atomic and molecular abundance indicators.  The mean difference in equivalent width and oscillator strength for the lines in common is negligible, and would yield an offset of less than 0.01 dex if applied under the assumption that all of the lines are on the linear part of the curve of growth.  We illustrate the differences in Figure 4, defined in the sense (Biemont- AGSAB05).   In this figure we have corrected the Biemont abundances to the AGSAB05 equivalent width and oscillator strengths.  The differences are significant even for weak lines, suggesting that differences in the classical line broadening are probably not responsible.  A similar, but smaller, effect is present in the forbidden line.  \citet{pla2002} inferred a HM abundance of 8.48, which would also be obtained from \citet{sh1990} when a blending correction is made to the equivalent width. AGSAB05 could not trace a comparable difference (0.06 dex) relative to the earlier work of Lambert.  The only obvious source that we can derive is a note by Sturenburg \& Holweger that they corrected their atomic abundances for the fraction of C tied up in CO, which could be of the right order to explain the differences.  Until the origin of this discrepancy (which is not present for oxygen) is explained, we have to treat this as a systematic uncertainty in the atomic abundance scale.  Abundances derived under this scale are the last set of values in Table 3.

\subsubsection{Molecular Lines}

We consider the same four molecular indicators that were included in AGSAB05.  They chose to disregard one of them (CH electronic lines) in their derived mean abundances, on the grounds that they are located in a crowded portion of the spectrum and sensitive to the treatment of line broadening.  However, the formal errors in the CH electronic abundances are similar to those for the other molecular species, and as such we see no obvious reason to exclude them.  We do treat the CH (v,r) abundances as being more reliable, as they are based on many more lines than the other diagnostics.  We therefore assigned double weight to the CH values and single weight to both the C2 electronic and CH electronic values.  As for oxygen, the mean was derived by a weighted average of the carbon obtained with different molecular indicators, and the scatter of the individual line measurements for all diagnostics around the adopted mean was taken as a measure of the random error.  We did not include carbon (or oxygen) abundances derived from CO line studies, because there are complex correlated errors.  Had we included them, the net effect would have been to increase the molecular abundances relative to the atomic values. 

\subsubsection{Carbon Abundance and Error Analysis}

Our final inference concerning carbon depends on which atomic abundance scale is adopted.  If we take the low scale of AGSAB05, the 3D model atomic and molecular abundances (8.40, 8.42) are closer than those for HM74 (8.45, 8.55); 1D MARCS abundances are also consistent (8.40, 8.44).  We would estimate a mean value of 8.41-8.42, with a random error of 0.04 dex.  The HM74 average of 8.50 would be an effective $2 \sigma$ internal inconsistency, implying a 0.04 dex systematic uncertainty for a total abundance of 8.41 +/- 0.06.  Adopting the higher scale would give pairwise results of (8.44, 8.42), (8.50, 8.55), (8.44, 8.44); all three models are internally consistent within the errors, and a mean abundance would be 8.47 with a total error of 0.05 (0.04 random, 0.03 systematic).  We adopt the mean of these approaches (8.44), and estimate an error of 0.04 (random) and 0.04 (systematic) for a total of 0.06 dex when combined in quadrature.

\subsection{Nitrogen Abundance Indicators}

Our discussion of nitrogen is necessarily briefer than that of oxygen and carbon, largely because the published results are preliminary and incomplete.  Holweger (2001) derived a non-LTE [N/H] = 8.0 +/- 0.11, comparable to results in previous compilations of solar abundances from \citet{gs1998}.  The compilation of models in \citet{ags2005} yields atomic and molecular nitrogen abundance estimates of (7.85 +/-0.08, 7.73 +/- 0.05), (7.97 +/- 0.08, 7.95 +/- 0.05), (7.94 +/- 0.08, 7.82 +/- 0.05) for 3D, HM, and MARCS respectively.  The same correspondence between 3D and MARCS that was seen in oxygen is replicated in nitrogen, but the internal consistency in the HM model is higher than that in the other models.  The formal significance of the disagreement in the 3D models is under $2 \sigma$, however, so we cannot exclude the possibility that they may be consistent.  We therefore adopt the HM result as the central value (7.96 +/- 0.06), and treat the difference with the 3D result (7.78 +/- 0.06) as a $2 \sigma$ systematic error.  This yields a total uncertainty in [N/H] of 0.10 dex dominated by systematic uncertainties.

\subsection{The Solar Beryllium Abundance}

  There is an interesting linkage between the solar O and Be abundances.
  In stellar interiors Be is destroyed at modest temperatures (of order 3.5
  million K).  It can therefore be used as a diagnostic of mixing in stars,
  especially in conjunction with the more fragile light element Li
  \citep{mhp1997}.  Traditional model atmospheres studies \citep{cbm1975} yield a solar
  photospheric beryllium abundance roughly half of the meteoritic abundance.
  However, the only accessible Be feature is located in a crowded portion of
  the spectrum in the near UV, and the continuum opacity is uncertain in this
  regime (largely from the contribution of numerous weak iron lines).  Since
  the strength of a line is a function of the ratio of the line to the
  continuous opacity, a higher photospheric Be could be derived if the
  continuous opacity background was higher than that of the model. In an
  important paper, \citet{bb1998} pointed out that nearby OH
  lines could be used to test the continuous opacity close to beryllium.  They
  derived a UV OH lines that were too strong if they used the absolute oxygen
  abundance obtained from the IR OH lines, and interpreted this as evidence
  that the continuous opacity is underestimated in the spectral window
  relevant for Be.  Similar conclusions for 3D models were obtained by
  \citet{a2004}.  Following \citet{l2003}, we note that the
  uncertainties in the ad hoc corrections are substantial.  Asplund (2004)
  quotes photospheric and meteoritic abundance errors of 0.09 and 0.08 dex
  respectively, implying that his zero net photospheric depletion has a $1
  \sigma$ uncertainty of 0.12 dex.  Even if the Balachandran and Bell argument
  is entirely correct, the data sets a $2 \sigma$ limit of 0.24 dex on
  beryllium depletion and does not require that it be zero.  There is also the
  possibility of substantial NLTE corrections to the UV OH lines, which would
  reduce or even eliminate the requirement for a mechanism to reduce the
  strength of the lines.  We also note that the value of the oxygen used by
  \citet{bb1998} for the IR lines (8.91) in the HM74 model is
  larger than the value derived from other molecular and atomic indicators,
  and even slightly larger than the value from the (v,r) transitions in the
  same model from the work of Asplund and collaborators.  We contend that this
  promising approach still has substantial errors, including large
  uncertainties in the absolute solar oxygen abundance.  We therefore believe
  that the approach of \citet{l2003} is the best current picture of the
  degree of beryllium depletion in the Sun: namely, there is a substantial
  uncertainty in the degree of solar beryllium depletion, and that further
  work is required before powerful observational bounds can be used to constrain interiors calculations.

\section{Conclusions and Future Tests}

Our basic conclusion is simple: the difference between the solar CNO abundances as derived from model atmospheres and model interiors considerations is not statistically significant.  The systematic errors in photospheric abundance indicators will have to be reduced before a ``solar abundance problem'' can be established (or ruled out) with confidence. However, the disagreement between the solar thermal structure and that of the simulations would favor the higher abundance scale, and there is some recently published evidence to that effect.  If this is confirmed, it switches the nature of the problem from being a question of the correct abundance scale to a question of the uncertainties in numerical convection simulations. We begin with a synthesis and explanation of our findings.  We then divide our conclusions into two parts. We recommend steps to more firmly establish the photospheric abundance scale, and contend that accurate solar abundances require tests of the thermal structure of the models and the magnitude of non-LTE abundance corrections.  In our final subsection we then gather together evidence that the atmospheric abundance scale problem may be tied to the limited resolution in the convection simulations or errors in the underlying model atmosphere treatment.  The consequences for the solar beryllium abundance, which is a useful diagnostic of internal mixing, are also explored.

The two main justifications for the superiority of the 3D hydro atmospheres are the treatment of line broadening and the inclusion of granulation.  Both of these represent genuine improvements in the atmospheric physics.  However, neither of these effects is actually primarily responsible for the difference in the solar abundance scale.  Many of the abundance indicators are insensitive to the effective microturbulence.  If temperature fluctuations are imposed on a semi-empirical Holweger-Mueller atmosphere, the resulting granulation corrections are usually smaller than the 3D convection effects reported by Asplund and collaborators, and frequently opposite in sign \citep{h2001}.  The main driver behind the systematic reductions in abundance derived from the 3D models is a theoretically predicted change in the thermal structure, coupled with large assumed non-LTE corrections for atomic features.  Neither of these changes is directly supported by observational tests.  Instead, the argument for the superiority of the abundances derived from the newer model atmospheres is an indirect one, focused on the concordance of abundances derived from different indicators.

A consistent chain of logic emerges from the comprehensive studies of oxygen (AGSAK04) and carbon (AGSAB05). Classical LTE model atmospheres tend to yield internally consistent, and high, carbon and oxygen abundances for atomic and molecular indicators. The application of a different thermal structure in the 3D hydro atmospheres drastically reduces the abundances inferred from highly temperature sensitive molecular indicators, but has a smaller effect on atomic features. Large NLTE corrections are then applied to the abundances derived from permitted atomic features for both 1D and 3D models. 
The net result is that the abundance estimates from 1D models become internally inconsistent (atomic indicators yield lower abundances than molecular ones), while abundances derived from the 3D models are internally consistent.  The abundances derived from forbidden lines are insensitive to NLTE effects, but they are reduced in the newer generation of models by the inclusion of blending features.  As a secondary argument, the fits to individual indicators are argued to be superior in the 3D models when compared to the fits to individual indicators in the 1D models.  This approach is appealing on the surface, but when examined in detail the picture is decidedly more ambiguous.  If anything, the hints from the data would lean towards the opposite conclusion.

The abundances derived from forbidden lines have the smallest systematic errors, but errors in both the theoretical oscillator strengths and the treatment of blending features result in non-negligible random errors.  More to the point, the internal consistency of abundances derived from forbidden and molecular lines is actually similar in the 3D and 1D cases.  From Table 1, the forbidden and molecular oxygen abundances are (8.71, 8.64) for 3D and (8.77, 8.84) for the HM; the differences are identical.  Given the errors, neither discrepancy is statistically significant with high confidence.

The abundances reported for permitted atomic features in AGSAK04 and AGSAB05 are significantly lower for 1D models than the corresponding molecular abundances, while the reported 3D results are in agreement.  In the case of oxygen, this rests completely on the assignment of large NLTE corrections.  These corrections were obtained under the assumption that hydrogen collisions were unimportant.  Detailed studies of the response of the triplet to limb-darkening indicate that models including hydrogen collisions are favored, and the inferred NLTE corrections decrease.  As a result, the internal consistency of the oxygen indicators is comparable for the different classes of atmospheres.  Nitrogen is consistent for HM74 models and inconsistent (but at less than $2 \sigma$) for the 3D case.  In the case of carbon, the situation is made more complex by significant zero-point offsets between earlier studies of carbon abundances that are not explained.  Again, the assignment of larger NLTE corrections is uncertain (and, unlike the case of oxygen, not directly tested against limb-darkening data).  A clean distinction between models on the basis of consistency is not obtained.  However, the 3D models do yield different molecular and atomic abundances for both N and O, and might also do so for C.

One might then hope to find distinct differences in the quality of the fits to different molecular indicators.  The usual patterns, unfortunately, manifest themselves as simple zero-point shifts.  For every case where there are issues with the 1D models (e.g. small trends with excitation potential in the [O/H] derived from (v,r) OH transitions in the HM model) there are comparable or larger effects for the 3D models (e.g. substantial trends in the [O/H] derived from (r,r) OH transitions).  In a recent preprint, \citet{scott2006} examined CO indicators, and the resulting pattern is illustrative.  The 3D models yielded similar results for two of the three features studied, while the 1D models performed better in a different pair of indicators.  The $C^{12}/C^{13}$ ratio from the 1D models ranges from 69 to 84, while the same ratio for the 3D models ranges from 83 to 108.  These values should be contrasted with the expected terrestrial ratio of 89.  Scott et al. (2006) choose comparisons that favor the 3D models, while an advocate of the traditional models might reasonably stress the other cases.  In our view, the best choice of models is not clearly distinguishable from the CNO abundance studies.  We recommend caution when extrapolating these model results to other stars, where the differential effects can be even more drastic.

\subsection{Establishing the Absolute Photospheric Abundance Scale}

The single most important test that is required for atmospheres theory is a discriminant between the different proposed thermal structures of the solar atmosphere.  The recent paper by Ayres et al. (2006) makes an important contribution by making direct comparisons of solar data with the thermal properties of the simulations.  They present evidence that the solar center-to-limb variations in continuum flux are inconsistent with the predictions of the 3D hydro simulations.  They also note that the predicted magnitude of fluctuations in the upper atmosphere from the simulations appears to be larger than the observed pattern.  Ayres et al. then construct an empirical model of the atmosphere and derive a high oxygen abundance (8.85) from CO molecular features under the assumption of a fixed C/O ratio.  In retrospect this conclusion is not surprising.  The HM model is not a purely theoretical exercise; it was constructed to reproduce the mapping of the source function as a function of optical depth inferred from limb darkening studies of continuum flux and strong lines \citep[see also][]{arg1998}.  The relative trends we have inferred from atomic and molecular abundance indicators support the conclusions of Ayres et al., but the current errors make our evidence in this matter suggestive but not conclusive.  It would also be highly beneficial to repeat the HM exercise with the full 3D models as opposed to the restricted form of them that Ayres et al. (2006) had available to them.

Ultimately, the absolute accuracy of photospheric abundances is directly tied to the absolute accuracy of the thermal structure.  This suggests that an approach similar to that of \citet{sh2002} may be the optimal one.  In their paper they examined the impact of temperature fluctuations around an assumed mean empirical thermal structure, which in their case was the HM74 model.  Interestingly, the abundance corrections that they derive would act in the sense of increasing the concordance between abundance indicators.  Oxygen abundances from atomic indicators would be slightly increased; although they did not consider molecular features directly, the net effect would certainly have the same sign as that obtained from 3D hydro models, namely a decrease in the inferred abundance.  In such a differential approach, deviations between the mean structure of the simulations and the empirical data would be used as guidance concerning the underlying physics.  In contrast, the 3D model abundances assume that the ab initio profile is correct.  A similar approach could be employed for the velocity field that replaces the microturbulence and macroturbulence in traditional 1D atmospheres.  

A second ingredient that must be tested empirically, rather than by theoretical assertion, is the magnitude of NLTE corrections.  The available evidence suggests that NLTE corrections are in general small for the Sun, but for the level of precision required in the absolute abundance scale these small corrections are significant.  Studies of different spectral features yield different conclusions about the physical model employed in NLTE studies.  This implies that there are significant uncertainties in absolute theoretical calculations.  Fortunately, NLTE corrections can be constrained by the response of line strength to limb darkening in the Sun.  It should be possible to develop improved theoretical models with a sufficient database of information developed in this fashion.  One other stringent test of NLTE effects may be to focus on the species whose relative abundances can be reliably inferred from meteoritic data.  For example, NLTE effects may be significant for iron \citep{stb2001} but less so for Si \citep{w2001}.  \citet{h2001} noted that there may be a conflict between the photospheric and meteoritic Fe/Si ratio, albeit one of marginal significance.  A similar situation may exist for Na \citep{ags2005}.

Another tractable problem is the absolute error for the forbidden C and O lines.  In these cases, uncertainties in the line profiles and continuum levels should be included.  Better atomic data (such as oscillator strengths for both the lines and the blending features) would also be useful. 

The accuracy of the theoretically predicted turbulent velocity field as a function of optical depth should also be subjected to a more rigorous analysis.  \citet{scott2006} present evidence that the generation of simulations used for the abundance analysis yielded poor fits to the line bisectors of CO lines.  Higher resolution simulations gave better line profile fits, but for (unspecified) unrealistically high C/O abundances.  The higher resolution simulations were not employed in the CO abundance analysis in that paper.  It is worth keeping in mind that line profiles are integral quantities, and as a result the uniqueness of the solutions is not established by individual cases of good fits.  This is particularly true when the abundance itself is treated as a free parameter.  It would be extremely useful if future papers on abundances derived using numerical simulations illustrated individual line fits, as well as quantifying the actual impact of the ``effective microturbulence'' on the abundance estimates.

It is useful to separate out the impact of velocity broadening from the effect of granulation and temperature gradient changes.  This can be done by using the mean thermal structure and temperature fluctuations from the simulations and a more traditional micro/macroturbulence model to infer abundances, and comparing the results with the full 3D models.  \citet{scott2006} constructed such a test case (their 1DAV model), and found only small abundance offsets, of order 0.04 dex for oxygen derived from IR OH lines.  They also inferred carbon abundances from CO; in this case O was held fixed and the carbon was adjusted to fit different molecular indicators.  The deviations in the derived carbon abundances relative to the 3D case ranged from small (0.01 dex for the LE lines) to modest (0.06 dex for the weak $\Delta \nu = 1$ lines) to large (0.14 dex for the $\Delta \nu = 2$ lines). These deviations may explain the changes in excitation potential that \citet{ay2006} needed to obtain consistent abundances within a 1D framework. This exercise implies that the impact of the improved microphysics varies substantially for different indicators, and is worth quantifying across the board.  An alternate exercise (using the revised velocity field and relative temperature fluctuations while adopting a HM74 mean thermal structure) might also be illuminating.

\subsection{Uncertainties in Numerical Convection Simulations}

First-principles theoretical model atmosphere calculations have undeniable strengths.  The ability to naturally reproduce line widths and include granulation is a powerful addition to our ability to reliably interpret stellar and solar spectra.  The principal difficulty with such models is that errors in the input physics generate absolute errors in the inferred atmospheric structure that cannot be calibrated away in the absence of explicit free parameters.  This phenomenon is the major reason why numerical convection simulations have not replaced the simple mixing length theory in stellar interiors calculations.  Interiors models that can reproduce observed stellar radii are simply more useful for most purposes than models with a better physical treatment of convection that fail to do so.

Before the results from such models are adopted as the new abundance standard, it will be necessary to perform an extensive theoretical error analysis and to compare the models with the strongest observational constraints.  We believe that accurate solar abundance calculations must reproduce the observed solar thermal structure, and from the Ayres et al. (2006) paper the Asplund models employing numerical convection simulations appear to yield a temperature gradient steeper than the real Sun.  This could be caused by errors in the background (1D) stellar atmospheres treatment; for example, uncertainties in the equation of state and continuous opacities will induce absolute errors in the thermal structure.  An approach similar to that employed in interiors models would be useful for assessing the uncertainties in the thermal structure and abundance predictions, and this should be included in the error budget for abundances.

It is more likely, however, that the major error source in 3D hydro model atmospheres is related to uncertainties in the numerical convection simulations.
The approximations in hydro simulations of giant planet atmospheres are have been demonstrated to be strongly affected by the quality of the assumed physics \citep{eg2004}. \citet{zs2006} also provides a good summary of the uncertainties in the related problem of terrestrial and solar dynamo models.  Another phenomenon that could be related is the issue of convective overshooting below surface convection zones.  Numerical simulations have tended to favor extensive overshooting, and the early models had a substantial nearly adiabatic overshoot region, in conflict with the stringent limits set by seismology (less than $ 0.05 H_p$).  More recent 3D \citet{bct2002} and 2D \citep{rg2005,rg2006} calculations found that the filling factor for plumes is smaller than previously thought, which led to an overestimate in earlier models of the changes induced by overshooting in the thermal structure.  The newer simulations predict strongly subadiabatic overshooting (effectively, overmixing without changing the thermal structure), which is consistent with the seismic limits.  However, they still produce a substantial mixed region below the surface convection zone of order $0.4 H_p$.  Since even a small overmixing of $0.05 H_p$ drastically increases pre-MS lithium depletion \citep{mhp1997}, which is already too efficient relative to stellar data \citep{ptc2002}, it is likely that even this reduced overshooting is too large to be compatible with stellar constraints.  We argue that there is a common pattern in both "undershooting" and "overshooting" above and below convective regions.  In both cases, the numerical simulations may be overestimating the degree of mixing and the impact on the thermal structure of convection outside the formal bounds set by the Schwartzschild criterion.

There are two plausible error sources that should be investigated.  The treatment of heat transfer in the atmosphere convection simulations is necessarily simplified, and this may be leading to an artificial inhibition in energy transport between turbulent cells projected into the radiative atmosphere and their surroundings.  Resolution effects, however, may be even more important.  Even the highest resolution simulations available today are many orders of magnitude away from being able to reproduce the characteristic Reynolds numbers in the Sun.  Scott et al. (2006) found significant changes in line bisectors for the outer layers of their solar model when they increased their resolution, and these changes were in the sense of reducing the temperature contrast in the upper atmosphere and improving the shape of the bisectors relative to data.  Numerical tests with substantially increased resolution may shed some interesting light on the sensitivity of the predictions to the underlying numerics; 2D convection simulations may be useful in this regard.  We are optimistic that the net effect of such testing will be a greatly improved understanding of the strengths and weaknesses of theoretical atmospheres models, just as we are confident that the net result of the solar abundance controversy will be a far more secure knowledge of stellar abundances. 

\acknowledgments 

We would like to thank Martin Asplund for providing tables of the abundances derived from molecular oxygen abundance indicators.  We would also like to thank Don Terndrup, Jennifer Johnson, Andreas Korn, Chris Sneden, and Hans Ludwig for helpful discussions on stellar abundance determinations.  FD would like to thank Claude Zeippen for discussions on the uncertainties in the atomic data.



\begin{table}
\centering
\begin{tabular}{crrrrrr}
\hline
\noalign{\smallskip}
 Indicator & Mean & Error & Mean & Error & Mean & error \\
\noalign{\smallskip}
            \noalign{\smallskip}
\hline
\noalign{\smallskip} 
 &  3D  &    &  HM04  &    &  M04  &    \\
 $[$O I$]$  &  8.680  &  0.010  &  8.760  &  0.010  &  8.720  &  0.010  \\
 O I  &  8.639  &  0.021  &  8.644  &  0.065  &  8.716  &  0.031  \\
 OH (v,r)  &  8.610  &  0.030  &  8.870  &  0.030  &  8.740  &  0.030  \\
 OH(r,r)  &  8.650  &  0.020  &  8.820  &  0.010  &  8.830  &  0.030  \\
{\bf Atomic } &	{\bf 8.660 } &	{\bf n/a } &	{\bf 8.702 } &	{\bf n/a } &	{\bf 8.718 } &	{\bf n/a } \\
{\bf Molecular } &	{\bf 8.630 } &	{\bf n/a } &	{\bf 8.845 } &	{\bf n/a } &	{\bf 8.785 } &	{\bf n/a } \\
{\bf Atom-Mol } &	{\bf 0.030 } &	{\bf n/a } &	{\bf -0.143 } &	{\bf n/a } &	{\bf -0.067 } &	{\bf n/a } \\
\hline
 &  Rev3D  &    &  RevHM  &    &  RevMARCS  &    \\
 $[$O I$]$  &  8.700  &  0.055  &  8.780  &  0.055  &  8.740  &  0.055  \\
 O I  &  8.721  &  0.074  &  8.742  &  0.065  &  8.794  &  0.064  \\
 OH (v,r)  &  8.612  &  0.028  &  8.865  &  0.030  &  8.740  &  0.027  \\
 OH(r,r)  &  8.674  &  0.037  &  8.826  &  0.017  &  8.850  &  0.050  \\
{\bf Atomic } &	{\bf 8.707 } &	{\bf 0.044 } &	{\bf 8.764 } &	{\bf 0.042 } &	{\bf 8.763 } &	{\bf 0.042 } \\
{\bf Molecular } &	{\bf 8.635 } &	{\bf 0.048 } &	{\bf 8.835 } &	{\bf 0.031 } &	{\bf 8.765 } &	{\bf 0.082 } \\
{\bf Atom-Mol } &	{\bf 0.073 } &	{\bf 0.065 } &	{\bf -0.071 } &	{\bf 0.052 } &	{\bf -0.002 } &	{\bf 0.092 } \\

\noalign{\smallskip}
\hline\\
\end{tabular}
\vskip -0.5cm
\caption {Comparison between the O abundances derived from several indicators. The upper part of the tables
recals the value obtained by \citet{ags2005} while the bottom part summarize the our reanalysis (see text for details).}

\end{table}

\begin{table}
\centering
\begin{tabular}{clcrrr}
\hline
\noalign{\smallskip}
Indicator	&NLTE triplet correction&Wavelength (\AA)&	3D	&	HM04	&	M04\\
\noalign{\smallskip}
            \noalign{\smallskip}
\hline
\noalign{\smallskip}	
 O I	&0.000	&	6158.1	&	8.650	&	8.820	&	8.800	\\
 O I	&	&	7771.9	&	8.910	&	8.890	&	8.950	\\
 O I	&	&	7774.2	&	8.890	&	8.870	&	8.940	\\
 O I	&	&	7775.4	&	8.860	&	8.880	&	8.910	\\
 O I	&	&	8446.7	&	8.680	&	8.690	&	8.750	\\
 O I	&	&	9266.01	&	8.730	&	8.790	&	8.770	\\
{\bf Mean}&	& &	{\bf 8.803 } & {\bf 8.840 } & {\bf 8.872 } \\
{\bf Dispersion }& & & {\bf 0.105 } & {\bf 0.071 } & {\bf 0.084 } \\
 O I	&0.060	&	6158.1	&	8.642	&	8.806	&	8.792	\\
 O I	&	&	7771.9	&	8.836	&	8.811	&	8.885	\\
 O I	&	&	7774.2	&	8.825	&	8.796	&	8.877	\\
 O I	&	&	7775.4	&	8.805	&	8.809	&	8.855	\\
 O I	&	&	8446.7	&	8.658	&	8.660	&	8.728	\\
 O I	&	&	9266.01	&	8.708	&	8.760	&	8.748	\\
{\bf Mean}& &	&	{\bf 8.758 } & {\bf 8.787 } & {\bf 8.829 }	\\
{\bf Dispersion }& &	& {\bf 0.080 } & {\bf 0.055 } & {\bf 0.064 }	\\
 O I	&0.110	&	6158.1	&	8.635	&	8.795	&	8.785	\\
 O I	&	&	7771.9	&	8.775	&	8.745	&	8.830	\\
 O I	&	&	7774.2	&	8.770	&	8.735	&	8.825	\\
 O I	&	&	7775.4	&	8.760	&	8.750	&	8.810	\\
 O I	&	&	8446.7	&	8.640	&	8.635	&	8.710	\\
 O I	&	&	9266.01	&	8.690	&	8.735	&	8.730	\\
{\bf Mean} &	&	& {\bf 8.721 }	& {\bf 8.742 } & {\bf 8.794 } \\
{\bf Dispersion	}&	&	& {\bf 0.060 } & {\bf 0.049 } & {\bf 0.048 }	\\
 O I	&0.160	&	6158.1	&	8.628	&	8.784	&	8.778	\\
 O I	&	&	7771.9	&	8.714	&	8.679	&	8.775	\\
 O I	&	&	7774.2	&	8.715	&	8.674	&	8.773	\\
 O I	&	&	7775.4	&	8.715	&	8.691	&	8.765	\\
 O I	&	&	8446.7	&	8.622	&	8.610	&	8.692	\\
 O I	&	&	9266.01	&	8.672	&	8.710	&	8.712	\\
{\bf Mean}&	&	&	{\bf 8.684 } & {\bf 8.697 } & {\bf 8.759 }	\\
{\bf Dispersion}&	&	& {\bf 0.041 } & {\bf 0.052 } &	{\bf 0.035 }	\\
 O I	& variable &	6158.1	&	8.620	&	8.770	&	8.770	\\
 O I	&	&	7771.9	&	8.640	&	8.600	&	8.710	\\
 O I	&	&	7774.2	&	8.650	&	8.600	&	8.710	\\
 O I	&	&	7775.4	&	8.660	&	8.620	&	8.710	\\
 O I	&	&	8446.7	&	8.600	&	8.580	&	8.670	\\
 O I	&	&	9266.01	&	8.650	&	8.680	&	8.690	\\
{\bf Mean } &	&	& {\bf 8.639 } & {\bf 8.644 } & {\bf 8.716 }	\\
{\bf Dispersion } & & & {\bf 0.021 } & {\bf 0.065 } & {\bf 0.031 }	\\

\noalign{\smallskip}
\hline\\
\end{tabular}
\vskip -0.8cm
\caption {Oxygen abundance derived from the O I lines using different NLTE corrections. 
The top panel represent the LTE values. The next 3 panels presents results normalized 
to NLTE corrections of 0.06, 0.11 and 0.16 dex respectively and the bottem panel 
recals the \citet{agsak2004} results (see text for details).}

\end{table}

\begin{table}
\centering
\begin{tabular}{crrrrrr}
\hline
\noalign{\smallskip}
 Indicator      &       Mean    &       Error   &       Mean    &       Error   &       Mean    &       error   \\
\noalign{\smallskip}
            \noalign{\smallskip}
\hline
\noalign{\smallskip}
	&	3D	&	&	HM	&	&	M	&\\
$[$C I $]$	&	8.390	&	0.040	&	8.450	&	0.040	&	8.400	&	0.040	\\
C I	&	8.360	&	0.030	&	8.390	&	0.030	&	8.350	&	0.030	\\
CH - vib-rot	&	8.380	&	0.040	&	8.530	&	0.040	&	8.420	&	0.040	\\
C2 electronic	&	8.440	&	0.030	&	8.530	&	0.030	&	8.460	&	0.030	\\
CH electronic	&	8.450	&	0.040	&	8.590	&	0.040	&	8.440	&	0.040	\\
{\bf Average }	&{\bf	8.394	}&{\bf 	0.036	}&{\bf	8.483	}&{\bf 	0.036	}&{\bf	8.405	}&{\bf 	0.036	}	\\
{\bf Deviation}	&{\bf	0.034	}&	&{\bf	0.068	}	&	&	{\bf	0.046	}	&		\\
{\bf ATM}	&{\bf	8.375	}&	&{\bf	8.420	}	&	&	{\bf	8.375	}	&		\\
{\bf MOL}	&{\bf	8.413	}&	&{\bf	8.545	}	&	&	{\bf	8.435	}	&		\\
{\bf ATM-MOL}	&{\bf	-0.038	}&	&{\bf	-0.125	}	&	&	{\bf	-0.060	}	&		\\
\hline 
Corrected Low Atomic Scale  &  3D  &    &  HM  &    &  M  &    \\
$[$C I $]$	&	8.390	&	0.040	&	8.450	&	0.040	&	8.400	&	0.040	\\
C I	&	8.406	&	0.042	&	8.441	&	0.035	&	8.398	&	0.031	\\
CH - vib-rot	&	8.380	&	0.051	&	8.530	&	0.039	&	8.420	&	0.039	\\
C2 electronic	&	8.440	&	0.037	&	8.530	&	0.030	&	8.460	&	0.033	\\
CH electronic	&	8.450	&	0.057	&	8.590	&	0.058	&	8.440	&	0.039	\\
{\bf Average} &	{\bf8.405}	&{\bf 	0.045	}&{\bf	8.495	}&{\bf 	0.039	}&{\bf	8.417	}&{\bf 	0.036}	\\
{\bf Deviation}	&{\bf	0.026	}	&	&	{\bf	0.049	}	&&{\bf	0.029	}	&		\\
{\bf ATM}&	{\bf	8.398	}	&	&	{\bf	8.445	}	&&{\bf	8.399	}	&		\\
{\bf MOL}&	{\bf	8.413	}	&	&	{\bf	8.545	}	&&{\bf	8.435	}	&		\\
{\bf ATM-MOL}&	{\bf	-0.014	}	&	&	{\bf	-0.100	}	&&{\bf	-0.036	}	&		\\
\hline
Corrected high Atomic Scale & 3D &  &HM  &  &M &  \\
$[$C I $]$	&	8.425	&	0.040	&	8.485	&	0.040	&	8.435	&	0.040	\\
C I	&	8.486	&	0.042	&	8.521	&	0.035	&	8.478	&	0.031	\\
CH - vib-rot&	8.380	&	0.051	&	8.530	&	0.039	&	8.420	&	0.039	\\
C2 electronic&	8.440	&	0.037	&	8.530	&	0.030	&	8.460	&	0.033	\\
CH electronic&	8.450	&	0.057	&	8.590	&	0.058	&	8.440	&	0.039	\\
{\bf Average }&	{\bf 8.434}&{\bf 0.045	}&{\bf	8.524	}&{\bf 	0.039}	&{\bf	8.446	}&{\bf 	0.036	}	\\
{\bf Deviation}&{\bf 0.044}	&&{\bf	0.021	}	&	&{\bf	0.026	}	&		\\
{\bf ATM}&{\bf 8.456}	&	&{\bf	8.503	}	&		&{\bf	8.457	}	&		\\
{\bf MOL}&{\bf 8.413}	&	&{\bf	8.545	}	&		&{\bf	8.435	}	&		\\
{\bf ATM-MOL}&	{\bf 0.043}	&&{\bf	-0.042	}	&	&{\bf	0.022	}	&		\\
													
\noalign{\smallskip}
\hline\\
\end{tabular}
\vskip -0.5cm
\caption {Comparison of the different C abundances derived from the indicators used by \citet{agsab2005}
when differents set of correction are applyed. The top panel corresponds to the results derived in \citet{agsab2005},
the middle panel correspond to the revised value when we adopt another NLTE correction that is smaller than those of
\citet{agsab2005} and the bottom panel includes a shifth of +0.035 dex and +0.08 dex in the forbiden and allowed lines 
respectively (see text for detail). }

\end{table}

\begin{figure}\plotone{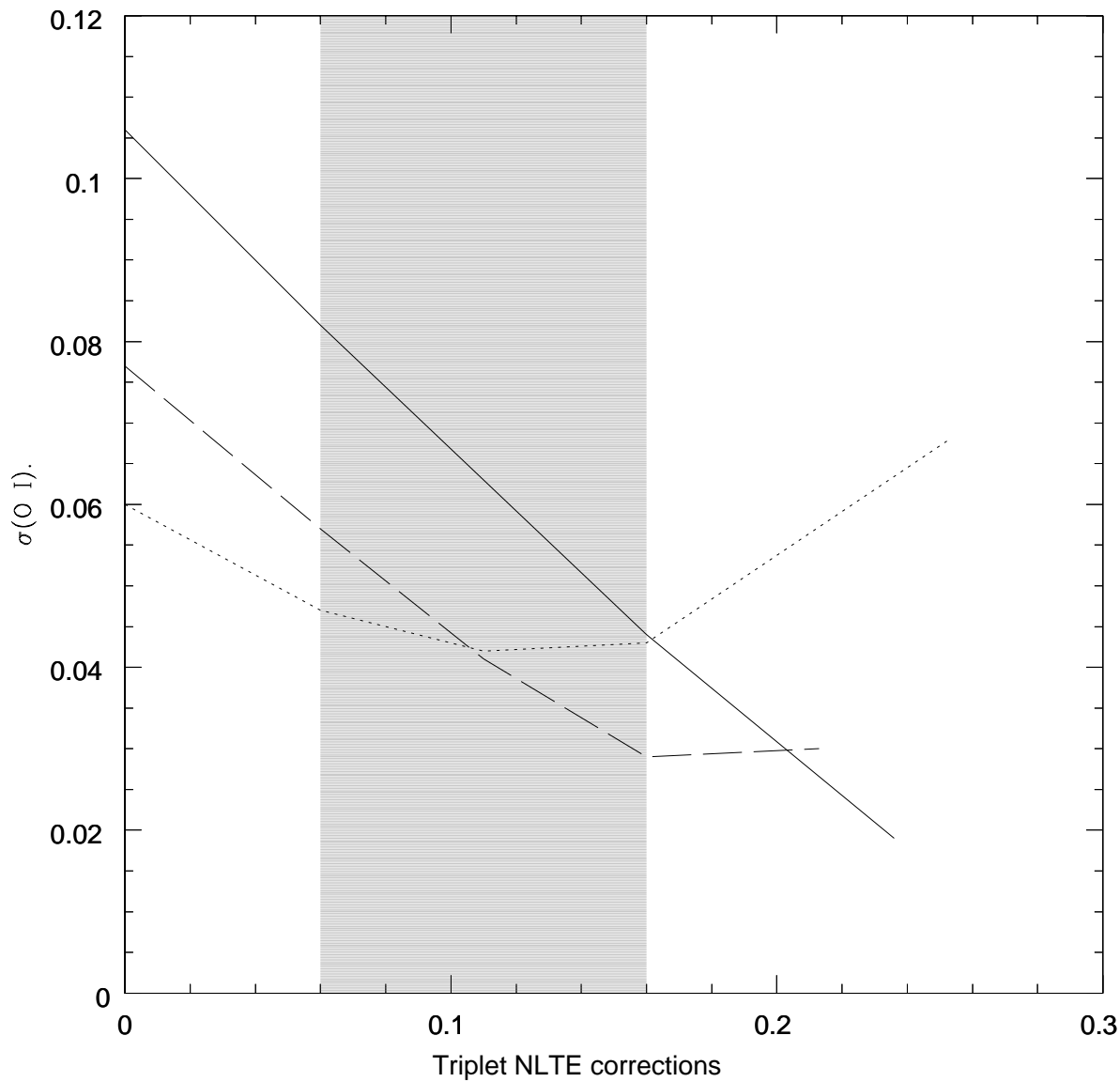}
\caption{The internal scatter in oxygen inferred from permitted 
atomic lines as a function of the
NLTE correction applied to the triplet.  The solid line is the 3D model,
the dotted line is the 1D HM74 model, and
the dash-dotted line is the 1D MARCS model.  AGSAK04 adopted the values on
the right side.  The band
denotes the limiting cases discussed in the text.  The 3D models are less
internally consistent than the 1D models
for NLTE corrections in this range.}
\end{figure}

\begin{figure}\plotone{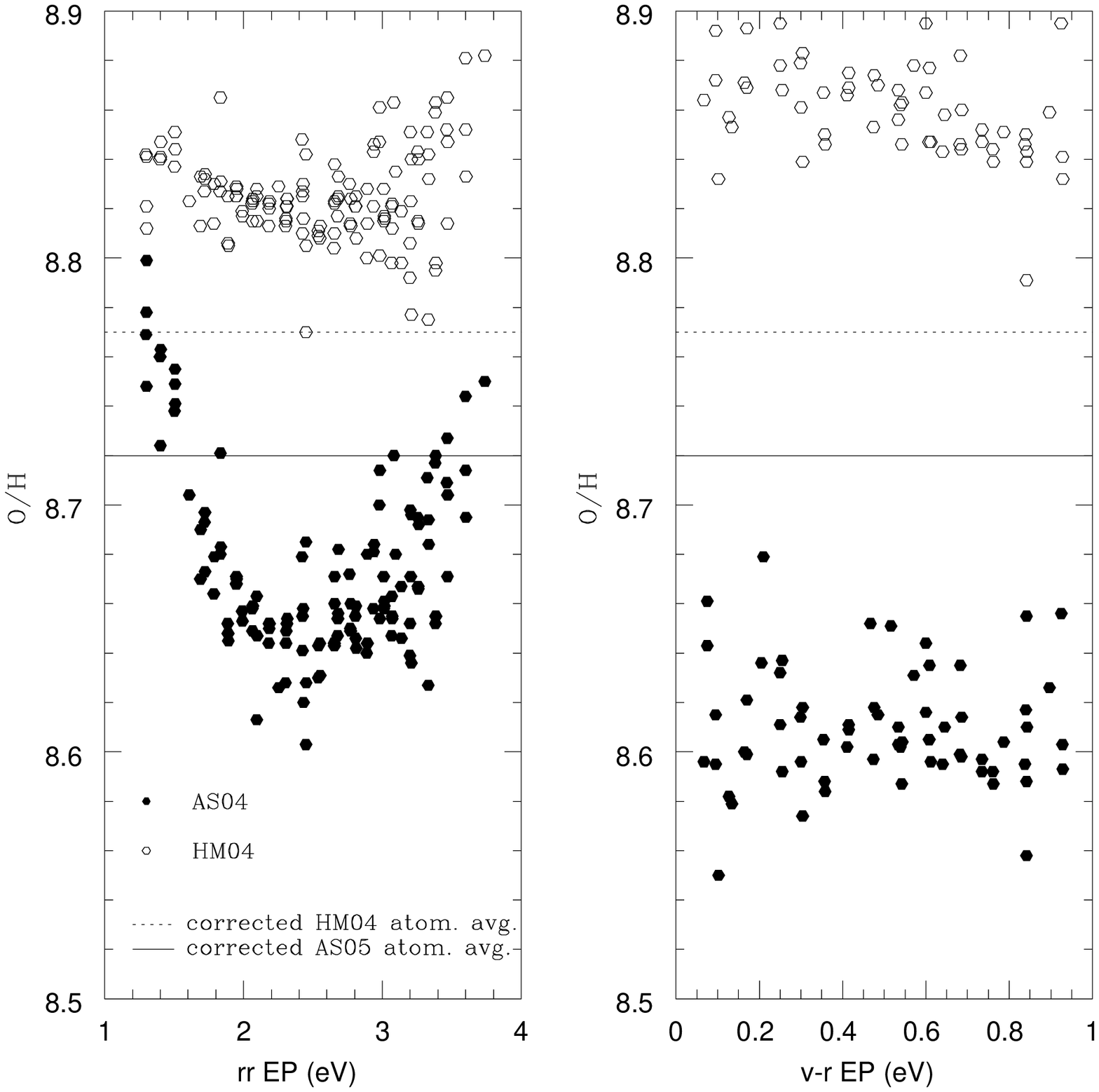}
\caption{The oxygen abundances derived from infrared OH lines in the 1D HM74 and 3D
AGSAK04 atmospheres
are compared as a function of excitation potential.  The left panel
represents (r,r) transitions,
while the right panel represents (v,r) transitions.  Results from the 3D
model atmospheres are solid circles;
results from the HM74 semi-empirical model atmosphere are open
circles.  The solid and dashed lines
indicate the mean abundances that would be derived from atomic features in
the 3D and HM74 models.
Comparable discrepancies between molecular and atomic abundances are
present for both classes of
models, and the 3D models exhibit striking trends with excitation potential
not seen in the HM74 case. }
\end{figure}

\begin{figure}\plotone{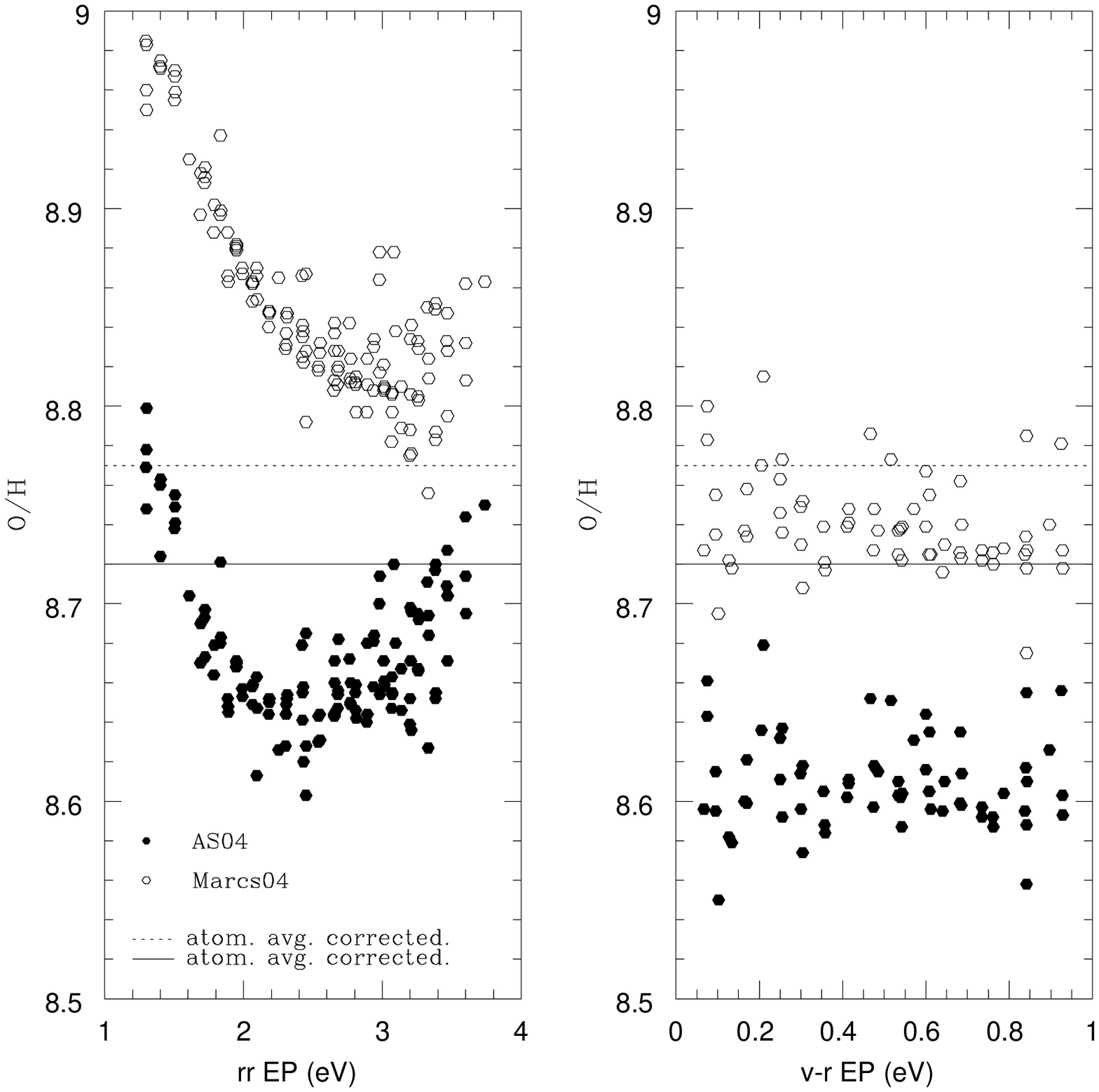}
\caption{The oxygen abundances derived from infrared OH lines in the 1D MARCS and
3D AGSAK04 atmospheres
are compared as a function of excitation potential.  The left panel
represents (r,r) transitions,
while the right panel represents (v,r) transitions.  Results from the 3D
model atmospheres are solid circles;
results from the MARCS theoretical 1D model atmosphere are open
circles.  The solid and dashed lines
indicate the mean abundances that would be derived from atomic features in
the 3D and MARCS models.
The MARCS model is more internally consistent than the 3D case, but similar
trends with excitation
potential are seen in both.  Since the underlying model atmosphere code is
similar, we conclude that the origin of
these trends is in the basic model atmospheres rather than being induced by
the treatment of convection. }
\end{figure}

\begin{figure}\plotone{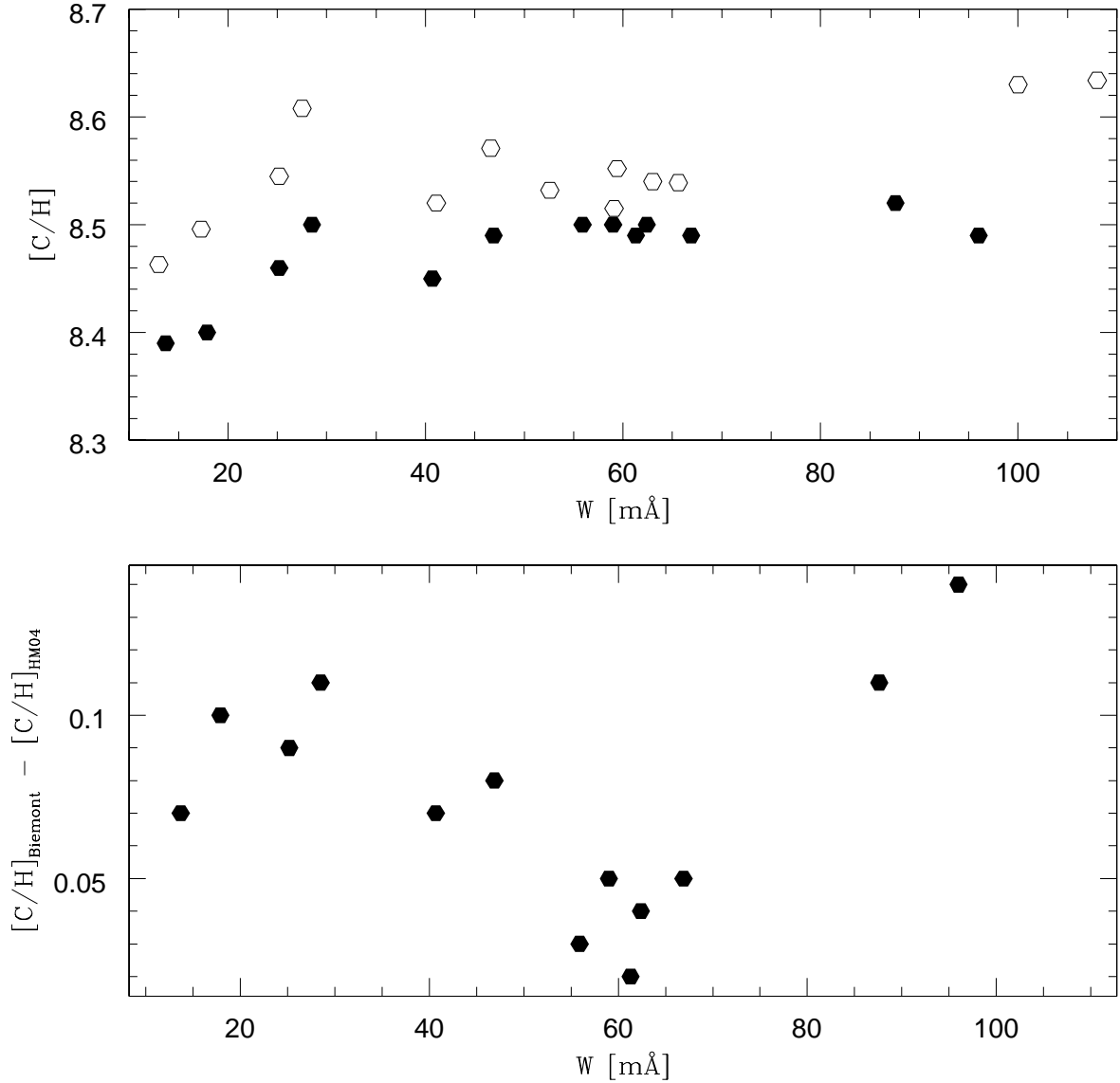}
\caption{Top panel: Comparison of the derived oxygen abundance from the
  atomic lines common to \citet{ags2005} (filled circles) and \citet{bhgv1993}
  (empty circles) as a function of the equivalent width. The data from
\citet{bhgv1993} have been corrected for the difference in log (gf) and in equivalent width.
bottom panel:
  Difference between these abundances as a function of equivalent width.}
\end{figure}


\begin{thebibliography}

\bibitem[Allende Prieto et al.(2004)]{pafb2004} Allende Prieto, C., Asplund, M., \& Fabiani Bendicho, P.\ 2004, \aap, 423, 1109 

\bibitem[Allende Prieto, Lambert \& Asplund (2001)]{pla2001} Allende Prieto,C., Lambert, D.~L., \& Asplund, M.\ 2001, \apjl, 556, L63

\bibitem[Allende Prieto, Lambert \& Asplund (2002)]{pla2002} Allende Prieto,C., Lambert, D.~L., \& Asplund, M.\ 2002, \apjl, 573, L137

\bibitem[Allende Prieto et al.(1998)]{arg1998} Allende Prieto, C., Ruiz Cobo, B., \& Garcia Lopez, R.~J.\ 1998, \apj, 502, 951 

\bibitem[Anders \& Grevesse(1989)]{ag1989} Anders, E., \& Grevesse, N.\ 1989, \gca, 53, 197 

\bibitem[Antia \& Basu(2006)]{ab2006} Antia, H.~M., \& Basu, S.\ 2006, \apj, in press (astro-ph/0603001)

\bibitem[Asplund(2004)]{a2004} Asplund, M.\ 2004, \aap, 417, 769 

\bibitem[Asplund et al.(2000a)]{asplund2000a} Asplund, M., Nordlund, {\AA}., Trampedach, R., \& Stein, R.~F.\ 2000, \aap, 359, 743 

\bibitem[Asplund et al.(2000b)]{asplund2000b} Asplund, M., Nordlund, {\AA}., Trampedach, R., Allende Prieto, C., \& Stein, R.~F.\ 2000, \aap, 359, 729 

\bibitem[Asplund(2000)]{a2000} Asplund, M.\ 2000, \aap, 359, 755 

\bibitem[Asplund, Grevesse \& Sauval (2005)]{ags2005} Asplund, M., Grevesse, N., \& Sauval, A.~J.\ 2005, ASP Conf.~Ser.~336: Cosmic Abundances as Records of Stellar Evolution and Nucleosynthesis, 336, 25 

\bibitem[Asplund et al.(2005)]{agsak2005} Asplund, M., Grevesse, N., Sauval, A.~J., Allende Prieto, C., \& Kiselman, D.\ 2005, \aap, 435, 339 

\bibitem[Asplund et al.(2005)]{agsab2005} Asplund, M., Grevesse, N., Sauval, A.~J., Allende Prieto, C., \& Blomme, R.\ 2005, \aap, 431, 693 

\bibitem[Asplund et al.(2004)]{agsak2004} Asplund, M., Grevesse, N., Sauval,  A.~J., Allende Prieto, C., \& Kiselman, D.\ 2004, \aap, 417, 751 

\bibitem[Ayres et al.(2006)]{ay2006} Ayres, T.~R., Plymate, C. \& Keller, C.~U. 2006, \apjs, (in press) 


\bibitem[Balachandran \& Bell(1998)]{bb1998} Balachandran, 
S.~C., \& Bell, R.~A.\ 1998, \nat, 392, 791 

\bibitem[Basu \& Antia(2004)]{ba2004} Basu, S., \& Antia, 
H.~M.\ 2004, \apjl, 606, L85 

\bibitem[Biemont et al.(1993)]{bhgv1993} Biemont, E., Hibbert,A., Godefroid,  M., \& Vaeck, N.\ 1993, \apj, 412, 431

\bibitem[Biemont et al.(1991)]{bhgvf1991} Biemont, E., Hibbert, A., Godefroid, M., Vaeck, N., \& Fawcett, B.~C.\ 1991, \apj, 375, 818 

\bibitem[Brummell et al.(2002)]{bct2002} Brummell, N.~H., 
Clune, T.~L., \& Toomre, J.\ 2002, \apj, 570, 825 


\bibitem[Carlsson (1986)]{c1986} Carlsson,M., Uppsala Astronomical Observatory Repport No 33

\bibitem[Chmielewski et al.(1975)]{cbm1975} Chmielewski, Y., Brault, J.~W., \& Mueller, E.~A.\ 1975, \aap, 42, 37 

\bibitem[Christensen-Dalsgaard et al.(1995)]{cdmt1995} Christensen-Dalsgaard, J., Monteiro, M.~J.~P.~F.~G., \& Thompson, M.~J.\ 1995, \mnras, 276, 283 


\bibitem[Delahaye \& Pinsonneault(2006)]{dp2006} Delahaye, F., \& Pinsonneault, M.\ 2006, \apj, 647, in press 

\bibitem[Drawin (1968)]{d1968} Drawin, H.~W., Z. Phys, 211,404


\bibitem[Evonak \& Glatzmaier (2004)]{eg2004} Evonak, M. \& Glatzmaier, G. 2004 Geophysical and Astrophysical Fluid Dynamics 98, 241


\bibitem[Grevesse \& Noels(1993)]{gn1993} Grevesse, N., \& Noels, A.\ 1993, Physica Scripta Volume T, 47, 133 

\bibitem[Grevesse \& Sauval(1998)]{gs1998} Grevesse, N., \& Sauval, A.~J.\ 1998, Space Science Reviews, 85, 161 

\bibitem[Guzik et al.(2005)]{gwc2005} Guzik, J.~A., Watson,  L.~S., \& Cox, A.~N.\ 2005, \apj, 627, 1049 

\bibitem[Holweger(2001)]{h2001} Holweger, H.\ 2001, AIP Conf.~Proc.~598: Joint  SOHO/ACE workshop ''Solar and Galactic Composition'', 598, 23 

\bibitem[Holweger \& Mueller(1974)]{hm1974} Holweger, H., \& Mueller, E.~A.\ 1974, \solphys, 39, 19 


\bibitem[Johansson et al.(2003)]{j2003} Johansson, S., 
Litz{\'e}n, U., Lundberg, H., \& Zhang, Z.\ 2003, \apjl, 584, L107 


\bibitem[Kiselman(1993)]{k1993} Kiselman, D.\ 1993, \aap, 275, 269 


\bibitem[Lodders(2003)]{l2003} Lodders, K.\ 2003, \apj, 591, 1220 


\bibitem[Mel{\'e}ndez(2004)]{m2004} Mel{\'e}ndez, J.\ 2004, \apj, 615, 1042


\bibitem[Piau \& Turck-Chi{\`e}ze(2002)]{ptc2002} Piau, L., \& Turck-Chi{\`e}ze, S.\ 2002, \apj, 566, 419

\bibitem[Pinsonneault(1997)]{mhp1997} Pinsonneault, M.\ 1997, \araa, 35, 557 


\bibitem[Reetz (1998)]{r1998} Reetz, J. 1998, PhD thesis, Ludwig-Maximilians Univ.

\bibitem[Rogers \& Glatzmaier(2005)]{rg2005} Rogers, T.~M., \& Glatzmaier, G.~A.\ 2005, \apj, 620, 432

\bibitem[Rogers \& Glatzmaier(2006)]{rg2006} Rogers, T.~M., \& Glatzmaier, G.~A.\ 2006 submitted to ApJ - astro-ph/0601668


\bibitem[Schmelz et al.(2005)]{s2005} Schmelz, J.~T., Nasraoui, K., Roames, J.~K., Lippner, L.~A., \& Garst, J.~W.\ 2005, \apjl, 634, L197 

\bibitem[Scott et al.(2006)]{scott2006} Scott,P.~C., Asplund, M., Grevesse,  N.~,A., Sauval, J. \ 2006, \aap in press, astro-ph/0605116

\bibitem[Shchukina \& Trujillo Bueno(2001)]{stb2001} Shchukina, N., \&  Trujillo Bueno, J.\ 2001, \apj, 550, 970 

\bibitem[Steffen \& Holweger(2002)]{sh2002} Steffen, M., \& Holweger, H.\ 2002, \aap, 387, 258 

\bibitem[Stein \& Nordlund(1998)]{sn1998} Stein, R.~F., \& Nordlund, A.\ 1998, \apj, 499, 914 

\bibitem[Storey \& Zeippen(2000)]{sz2000} Storey, P.~J., \& Zeippen, C.~J.\ 2000, \mnras, 312, 813

\bibitem[Stuerenburg \& Holweger(1990)]{sh1990} Stuerenburg, S., \& Holweger, H.\ 1990, \aap, 237, 125 


\bibitem[Turck-Chi{\`e}ze et al.(2004)]{turck2004}  Turck-Chi{\`e}ze, S., Couvidat, S., Piau, L., Ferguson, J., Lambert, P., Ballot, J., Garc{\'{\i}}a, R.~A., \& Nghiem, P.\ 2004, Physical Review Letters, 93, 211102


\bibitem[Wedemeyer(2001)]{w2001} Wedemeyer, S.\ 2001, \aap, 373, 998 


\bibitem[Zhang \& Schubert(2006)]{zs2006} Zhang, K., \& Schubert, G.\ 2006, Reports of Progress in Physics, 69, 1581

\end{thebibliography}
\end{document}